\RequirePackage{fix-cm}
\documentclass[twocolumn,epjc3]{svjour3}  
\smartqed  
\RequirePackage{graphicx}
\RequirePackage{float}
\RequirePackage{hyperref}
\RequirePackage{amsmath}
\usepackage{mwe}
\RequirePackage{amssymb}
\RequirePackage{mathtools}   
\usepackage{orcidlink}
\usepackage{booktabs}
\usepackage{caption}
\usepackage{multirow}
\usepackage{dcolumn}
\usepackage[caption=false]{subfig}
\usepackage{float}   
\journalname{Eur. Phys. J. C}
\begin{document}

\title{Constraining effective equation of state in $f(Q,T)$ gravity}


\author{Simran Arora\thanksref{e1,addr1}
        \and
        Abhishek Parida\thanksref{e2,addr2}
        \and
        P.K. Sahoo\thanksref{e3,addr1}}

\thankstext{e1}{e-mail: moraes.phrs@gmail.com}
\thankstext{e2}{e-mail: abhishekparida22@gmail.com}
\thankstext{e3}{e-mail: pksahoo@hyderabad.bits-pilani.ac.in}


\institute{Department of Mathematics, Birla Institute of
Technology and Science-Pilani, Hyderabad Campus, Hyderabad-500078,
India \label{addr1}
           \and
           International College of Liberal Arts, Yamanashi Gakuin University, Yamanashi 400-0805 Japan. \label{addr2}
}

\date{Received: 26 March 2021 / Accepted: date}

\maketitle

\begin{abstract}
New high-precision observations are now possible to constrain different gravity theories. To examine the accelerated expansion of the Universe, we used the newly proposed $f(Q,T)$ gravity, where $Q$ is the non-metricity, and $T$ is the trace of the energy-momentum tensor. The investigation is carried out using a parameterized effective equation of state with two parameters, $m$ and $n$. We have also considered the linear form of $f(Q,T)= Q+bT$, where $b$ is constant. By constraining the model with the recently published 1048 Pantheon sample, we were able to find the best fitting values for the parameters $b$, $m$, and $n$. The model appears to be in good agreement with the observations. Finally, we analyzed the behavior of the deceleration parameter and equation of state parameter. The results support the feasibility of $f(Q,T)$ as a promising theory of gravity, illuminating a new direction towards explaining the Universe's dark sector.
\end{abstract}

\section{Introduction}\label{sec1}

Starting with the Big Bang and moving on to the birth of elements, dark matter, and anti-matter, some fundamental questions about the universe are now making their way into the scientific realm. However, there are now several solutions available to these issues. Around the same time, cosmological discoveries revealed that the universe seems to be expanding faster. This has been supported by observations of distant Type Ia Supernovae (SNe Ia) \cite{Riess,Perlmutter}, and it is one of the most intriguing questions in recent years. Furthermore, these results revealed that a mysterious component known as dark energy makes up roughly 70\% of the universe. 

General Relativity is a geometrical theory that has offered a definitive explanation of the observational evidence and has led to new ideas into the problems of space and time. Despite its extraordinary success, several recent observations have raised concerns regarding standard GR validity, which may still have certain limitations, particularly on cosmic scales and beyond the solar system.
The cosmological constant is the most obvious candidate for dark energy, but it has flaws like the coincidence problem and fine-tuning problem \cite{Copeland,Bull,Weinberg}. Rapid expansion can be associated with two aspects. The first method involves manipulating the energy-momentum tensor in Einstein's field equations, such as scalar fields (quintessence, phantom) \cite{Wang,Shinji,Caldwell}, exotic equations of state (Chaplygin gas) \cite{Bento,Alexander}, viscosity (bulk viscosity) \cite{Padmanabhan,Brevik}, and so forth. The second approach is to alter the geometry of spacetime in  Einstein's equations. This is often known as the modified theories of gravity.\\
Modified gravity is a prominent branch of modern cosmology that seeks to provide a unified explanation for the universe's early epoch while also accounting for its later accelerated expansion. One of the simplest modified gravity theories is the $f(R)$ \cite{Starobinsky,Buchdahl,Martino} gravity theory, which expresses gravitational motion in terms of an arbitrary function of $R$, the Ricci scalar.
However, the extra degree of freedom afforded by the metric formulation in $f(R)$ would conflict with observational data observed in the solar system. Another modified gravity based on a non-minimal interaction between matter and geometry has been proposed to pass the solar system level such as $f(R,T)$ theory \cite{Xing,Harko/2014,Moraes/2017,Harko/2011}, $f(R,G)$ theory \cite{Elizalde,Bamba/2010}.

The most successful gravity theory at this moment is General Relativity. Not only did Einstein and Hilbert's contributions to physics and cosmology have a significant impact on astronomy, but they also had a significant effect on mathematics. They employed Riemannian geometry, a geometric theory of gravity that is now regarded as one of modern science's key pillars. Till now, GR has three equivalent geometrical representations: the curvature representation; the teleparallel representation; and the symmetric teleparallel representation. Following the introduction of Riemannian geometry in 1918, Weyl \cite{Weyl} attempts to create a more general geometry that can be used to achieve a geometrical integration of electromagnetism and gravitation. As a result, a new unified teleparallel theory was proposed \cite{Hayashi/1979}. The torsion in the teleparallel theory is utilized to characterize gravitational phenomena and is referred to as the teleparallel equivalent of GR or $f(T)$ gravity \cite{Yi-Fu,Capozziello/2011}. Based on the third representation of GR, a new theory has been proposed \cite{Nester} named as symmetric teleparallel gravity, in which the gravitational interaction is geometrically defined using the non-metricity $Q$, representing the variability in the length of a vector during parallel transport. As a result, a new extended gravity called $f(Q)$ was developed. The cosmological consequences of the $f(Q)$ gravity principle with observational constraints have also been studied \cite{Jimenez/2018,Harko/2018,J.Lu,Lazkoz,Mandal}. It is observed that the cosmological evolution in $f(Q)$ gravity is similar to $\Lambda$CDM, but it shows deviation from $\Lambda$CDM at perturbation level\cite{Khyllep/2021}.\\
After that, Yixin et al. \cite{Yixin/2019,Yixin/2020} introduced a symmetric teleparallel gravity extension based on the coupling between non-metricity $Q$ and the trace of the energy-momentum tensor $T$, i.e., considering an arbitrary function $f(Q,T)$ in the gravitational action. Using some functional forms of $f(Q,T)$, some works in this modified gravity look at cosmological models and rapid expansion of the universe. It has recently been discovered that $f(Q, T)$ gravity dramatically alters the nature of tidal forces and the equation of motion in the Newtonian limit\cite{Yang/2021}. Also, it is mentioned that generalized metric theories can be compared among each other with experimental results and Newtonian gravity. Thus, the post-Newtonian limit is almost appropriate for comparing theoretical predictions with solar system observations. As a result, comparing Weyl type $f(Q, T)$ gravity predictions concerning tidal force changes with observable evidence from a wide range of astrophysical phenomena should provide some understanding of the fundamental characteristics of the gravitational interaction and its geometrical description. So, there is a motivation to study various theoretical, observational, and cosmological aspects in $f(Q, T)$ theory.
Arora et al. \cite{Arora} tried to see if $f(Q,T)$ gravity could solve the late-time acceleration conundrum without adding an extra form of dark energy from a cosmological perspective. Bhattacharjee et al. \cite{Bhattacharjee} investigated baryogenesis in $f(Q,T)$ gravity.\\
As it is well established, the EoS parameter defines the relationship between pressure and energy density. The EoS parameter is used to classify various phases of the universe's decelerated and accelerated expansion. The matter-dominated phase is represented by $\omega=0$. In the current accelerated period of evolution, $-1<\omega\leq 0$ represents the quintessence phase, $\omega=-1$ represents the cosmological constant ($\Lambda$CDM) model, and $\omega<-1$ is the phantom age. In this work, we use the reconstruction of an efficient equation of state parameter to understand the late-time acceleration in $f(Q,T)$ gravity. The effective equation of state is not influenced by the individual properties of the different components of the matter field. Model parameters constrained by observational results determine the current value of the effective equation of the state parameter. Observational cosmology is the science of using observations to research the structure, existence, and evolution of the universe. Microwave Background Radiation \cite{Eisenstein}, Type Ia Supernovae \cite{Riess,Perlmutter}, Cosmic Baryon Acoustic Oscillations \cite{Spergel}, Planck data, and other observational datasets are currently available for various measurements and are providing robust evidence for the universe's acceleration. As a consequence, we will use Pantheon datasets \cite{Scolnic} to constrain the model parameters. The recently proposed Supernovae Pantheon sample contains 1048 points covering the redshift range $0.01<z<2.26$. We use the MCMC ensemble sampler given by the emcee library.\\
The article is divided into various sections.  An overview of $f(Q,T)$ gravity is given in Section \ref{sec2}. In section \ref{sec3}, we discussed the cosmological model and the parameterized equation of state and obtained an expression of the Hubble parameter. The brief discussion on observational data used to constrain the model parameters is given in Section \ref{sec4}. Section \ref{sec5} includes the behavior of cosmological parameters such as deceleration parameter and EoS parameter. The last section \ref{sec6} contains the concluding remarks. 

\section{Overview of $f(Q,T)$ Gravity}\label{sec2}

The action which is used to define $f(Q,T)$ gravity read as \cite{Yixin/2019},

\begin{equation}  \label{1}
S=\int \left( \frac{1}{16\pi} f(Q,T) + L_{m}\right )\sqrt{-g} d^4x,
\end{equation}
where $f$ is an arbitrary function that couples the non-metricity $Q$ and the trace of the energy-momentum tensor $T$. Also, $L_{m}$ represents the matter Lagrangian and $g = det(g_{\mu \nu})$. We also define the nonmetricity $Q$ as \cite{Jimenez/2018}

\begin{equation}  \label{2}
Q\equiv -g^{\mu \nu}(L^{ \beta}_{\,\,\, \alpha \mu}L^{\alpha}_{\,\,\, \nu \beta}-L^{\beta}_{\,\,\, \alpha \beta}L^{\alpha}_{\,\,\, \mu \nu}),
\end{equation}
where $L^{\beta}_{\,\,\,  \alpha\gamma}$ is the disformation tensor written as, 
\begin{equation}  \label{3}
L^{\beta}_{\alpha\gamma}=-\frac{1}{2}g^{\beta\eta}(\nabla_{\gamma}g_{\alpha \eta}+\nabla_{\alpha}g_{\eta\gamma}-\nabla_{\eta}g_{\alpha \gamma}).
\end{equation}

The nonmetricity tensor is represented by
\begin{equation}
\label{4}
Q_{\gamma\mu\nu}=\nabla_{\gamma}g_{\mu\nu},
\end{equation}
with trace of the non-metricity tensor given as
\begin{equation}
\label{5}
Q_{\beta}= g^{\mu \nu}Q_{\beta \mu \nu}  \qquad \widetilde{Q}_{\beta}= g^{\mu \nu}Q_{\mu \beta \nu}.
\end{equation}
We can also define a superpotential or the non-metricity conjugate as
\begin{equation}
\label{6}
P^{\beta}_{\,\,\, \mu \nu}= -\frac{1}{2} L^{\beta}_{\,\,\, \mu \nu}+ \frac{1}{4} (Q^{\beta}- \widetilde{Q}^{\beta})g_{\mu \nu} - \frac{1}{4} \delta^{\beta}_{(\mu}Q_{\nu)}.
\end{equation}
giving the nonmetricity scalar as \cite{Jimenez/2018}
\begin{equation}
\label{7}
Q=-Q_{\beta\mu\nu}P^{\beta\mu\nu}\,.
\end{equation}

In addition the energy-momentum tensor is known to be defined as
\begin{equation}  \label{8}
T_{\mu \nu}= -\frac{2}{\sqrt{-g}} \dfrac{\delta(\sqrt{-g}L_{m})}{\delta g^{\mu \nu}}
\end{equation} 
and 
\begin{equation}   \label{9}
\Theta_{\mu \nu}= g^{\alpha \beta} \frac{\delta T_{\alpha \beta}}{\delta g^{\mu \nu}}.
\end{equation}

Also, the variation of energy-momentum tensor with respect to the metric tensor is such that
\begin{equation}  \label{10}
\frac{\delta\,g^{\,\mu\nu}\,T_{\,\mu\nu}}{\delta\,g^{\,\alpha\,\beta}}= T_{\,\alpha\beta}+\Theta_{\,\alpha\,\beta}\,.
\end{equation}

Thus, taking the variation of action \eqref{1} with respect to the metric and equating it to zero, we get the following field equations:
\begin{multline}  \label{11}
-\frac{2}{\sqrt{-g}}\nabla_{\beta}(f_{Q}\sqrt{-g} P^{\beta}_{\,\,\,\, \mu \nu}-\frac{1}{2}f g_{\mu \nu}+ f_{T}(T_{\mu \nu}+\Theta_{\mu \nu})\\
-f_{Q}(P_{\mu \beta \alpha}Q_{\nu}^{\,\,\, \beta \alpha}-2Q^{\beta \alpha}_{\, \, \, \mu}P_{\beta \alpha\nu})= 8\pi T_{\mu \nu}.
\end{multline}
where $f_{Q}= \dfrac{df}{dQ}$ and $f_{T}= \dfrac{df}{dT}$.

Now, let us assume that the Universe is described by the homogeneous, isotropic and spatially flat FLRW metric given by, 
\begin{equation}  \label{12}
ds^{2}= -dt^{2}+ a^{2}(t)\delta_{ij}dx^{i}dx^{j},
\end{equation}
where $a(t)$ is the scale factor of the Universe. Moreover, the matter content of the Universe is assumed to be consisting of a perfect fluid, for which energy-momentum tensor is $T^{\mu}_{\,\,\, \nu} = diag(-\rho, p, p, p)$. Also, the nonmetricity function $Q$ for such a metric is calculated and obtained as $Q=6H^{2}$.

Using the metric and the field equation \eqref{11}, the generalized Friedmann equations are obtained as,
\begin{equation} \label{13}
8 \pi \rho= \frac{f}{2} -6 F H^{2}- \frac{2 \widetilde{G}}{1+\widetilde{G}}(\dot{F} H+ F\dot{H}),
\end{equation}

\begin{equation}\label{14}
8 \pi p= -\frac{f}{2} + 6 F H^{2} + 2(\dot{F}H+F \dot{H}).
\end{equation}
Here ($\cdot$)dot represents a derivative with respect to time, besides $F= f_{Q}$, and $8 \pi \widetilde{G}=f_{T}$ denote differentiation with respect to $Q$, and $T$, respectively.

Using the above two equations \eqref{13} and \eqref{14}, we can define the equations similar to the form of standard general relativity(GR),
\begin{equation}  
\label{15}
3H^2=8\pi \rho_{eff}= \frac{f}{4F}-\frac{4\pi}{F} \left[(1+ \widetilde{G})\rho+\widetilde{G} p\right],
\end{equation}
and 
\begin{multline}  \label{16}
2\dot{H}+ 3H^2=-8 \pi p_{eff}= \frac{f}{4F}-\dfrac{2\dot{F}H}{F}+\\
\frac{4\pi}{F} \left[(1+ \widetilde{G})\rho+(2+\widetilde{G}) p\right].
\end{multline} 
Moreover, $\rho_{eff}$, and $p_{eff}$ are the effective density and effective pressure respectively.

\section{Cosmological model and Equation of State}\label{sec3}

We consider the simplest functional form $f(Q,T)= Q+bT$, where $b$ is a constant. So, we get $F= f_{Q}= 1$ and $8\pi \widetilde{G}= b$.
Now, Solving for p and $\rho$ from Eqs.\eqref{13} and \eqref{14}. We obtained the effective or total equation of state parameter $\omega= \frac{p}{\rho}$ as
\begin{equation}
\label{17}
\omega= \frac{3 H^{2}(8\pi + b)+ \dot{H}(16 \pi+ 3b)}{ b \dot{H} - 3 H^{2}(8\pi+ b)}.
\end{equation}

Also, using the relation $\frac{a_{0}}{a}= 1+z$, we can define the relation for t and z as mentioned below.
\begin{equation}
\label{18}
\frac{d}{dt}= \frac{dz}{dt}\frac{d}{dz}= -(1+z) H(z)\frac{d}{dz},
\end{equation}
Normalizing the present value of scale factor to be $a_{0}= a(0)=1$. The Hubble parameter can be written in the form of 
\begin{equation}
\label{19}
\dot{H}= -(1+z) H(z)\frac{dH}{dz}.
\end{equation}

To solve equation \eqref{17} for $H$, we need one more alternate equation. So, we assume a well-motivated parametric form of equation of state parameter as a function of redshift $z$ \cite{Ankan/2016}, 
\begin{equation}
\label{20}
\omega= -\frac{1}{1+ m(1+z)^{n}},
\end{equation}
where $m$ and $n$ are model parameters. A. Mukherjee \cite{Ankan/2016} has described the behavior of this considered equation of state parameter. At the epoch of recent acceleration, it has a negative value of less than $-\frac{1}{3}$. For positive values of the model parameters $m$ and $n$, the value of $\omega$ tends to zero at a high redshift $z$ and depends on the model parameter at $z = 0$. The functional model of the effective equation of state assumed for the current reconstruction conveniently accommodates these two phases of evolution. A positive model parameter often sets a lower bound on the value of $\omega$ and keeps it in the non-phantom regime.

Using Eqs. \eqref{17}, \eqref{19} and \eqref{20}, we obtained the Hubble parameter $H$ in terms of redshift $z$.
\begin{equation}\label{21}
H(z)= H_{0} \left( \frac{b+(16\pi+ 3b)(1+m(1+z)^{n})}{b+(16\pi+3b)(1+m)}\right) ^{l},
\end{equation}

where $l= \frac{3(8\pi+b)}{n(16 \pi+3 b)}$, $H_{0}$ is the Hubble value at $z=0$.

\section{Observational data}\label{sec4}

This section describes the observational dataset used to constrain the model parameters $b$, $m$, and $n$ after obtaining the solutions for our model. To explore the parameter space, we use the MCMC sampling technique and mainly employ Python's emcee \cite{Mackey} library. For estimating the parameters, one need not compute the evidence, which is a normalizing constant. Instead, the prior and likelihood are sufficient to determine the posterior distributions of the parameters.\\
We used the recent Pantheon dataset for our work; it consists of 1048 Supernova Type Ia experiment results, discovered by the Pan-STARRS1(PS1) Medium Deep Survey, the Low-z, SDSS, SNLS, and HST surveys \cite{Scolnic,Chang}, in the redshift range $z \in (0.01, 2.26)$.

The chi-square ($\chi^{2}$) function, which relates the predictions of the model and SNe Ia observations read as,
\begin{equation} \label{23}
 \chi^{2}_{SN}(b,m,n )= \sum_{\mathfrak{D}} \frac{[\mu^{th}(b, m, n , z_{i})-\mu^{obs}(z_{i})]^{2}}{\sigma^{2}_{\mu(z_{i})}},
\end{equation}
where $\mathfrak{D}$ is the Pantheon samples with 1048 points, $\sigma^{2}_{\mu(z_{i})}$ is the standard error in the observed value, $\mu_{i}^{th}$ represents the theoretical value of distance modulus and $\mu_{i}^{obs}$ is the observed value of distance modulus.
\begin{equation} \label{24}
\mu_{i}^{th}= \mu(D_{l})= m -M = 5 log_{10}D_{l}(z)+ \mu_{0},
\end{equation}
where $m$ and $M$ are the apparent and absolute magnitudes with
\begin{equation}
\mu_{0}= 5 log(H_{0}^{-1}/Mpc)+25.
\end{equation}
as the marginalized nuisance parameter.
The formula for $D_{l}$ is given by
 \begin{equation} \label{25}
D_{L}(z)= (1+z) c \int_{0}^{z} \frac{dz'}{H(z')}.
\end{equation}

The 2-$\sigma$ bounds for the parameters from our analysis are $b = 0.2^{+2.7}_{-2.9} \ , m = 0.47^{+0.27}_{-0.21},\ n = 3.2^{+1.8}_{-2.0}$. As the functional form $f$ contains $b$ as a model parameter and $m$, $n$ are the parameters in the parametric functional form of the equation of state $\omega$, refer to Eq \eqref{17}.

\begin{figure*}[htbp]
\centering
\includegraphics[scale=0.6]{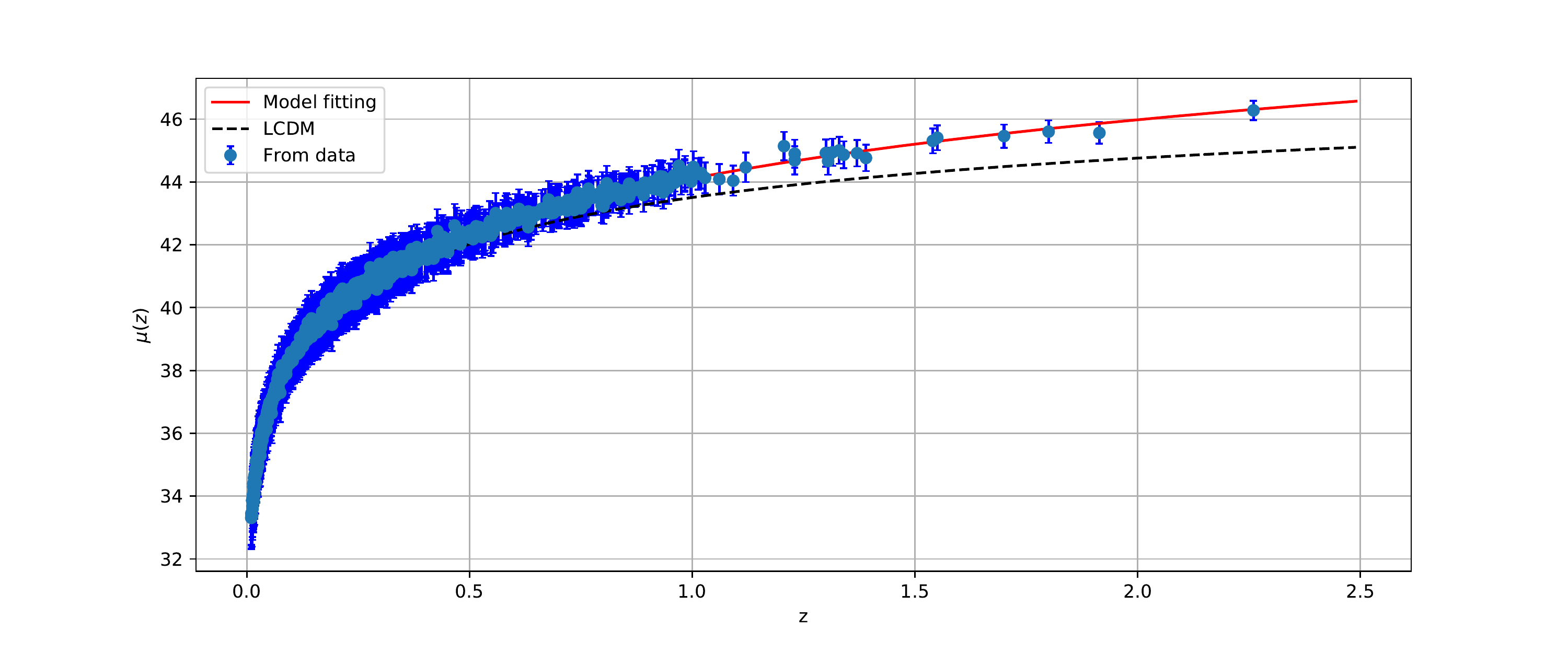}
\centering \caption{The plot of distance modulus $\mu(z)$ vs. redshift $z$ for our model shown in red line and $\Lambda$CDM in black dotted line which shows nice fit to the 1048 points of the Pantheon datasets shown with it's error bars.}
 \label{Fig-muz}
\end{figure*}

Fig. \ref{Fig-muz} shows a comparison between our model and the widely-accepted $\Lambda$CDM model in cosmology; we consider $\Omega_{m_{0}}= 0.3$, $\Omega_{\Lambda_{0}} = 0.7$ and $H_{0}= 67.4\pm 0.5km s^{-1}Mpc^{-1}$ \cite{Planck} for the plot. The figure also includes the Pantheon experimental results, 1048 data points along with their error, and allows for a clear comparison between the two models.

\begin{figure*}[htbp]
\centering
\includegraphics[scale=1.2]{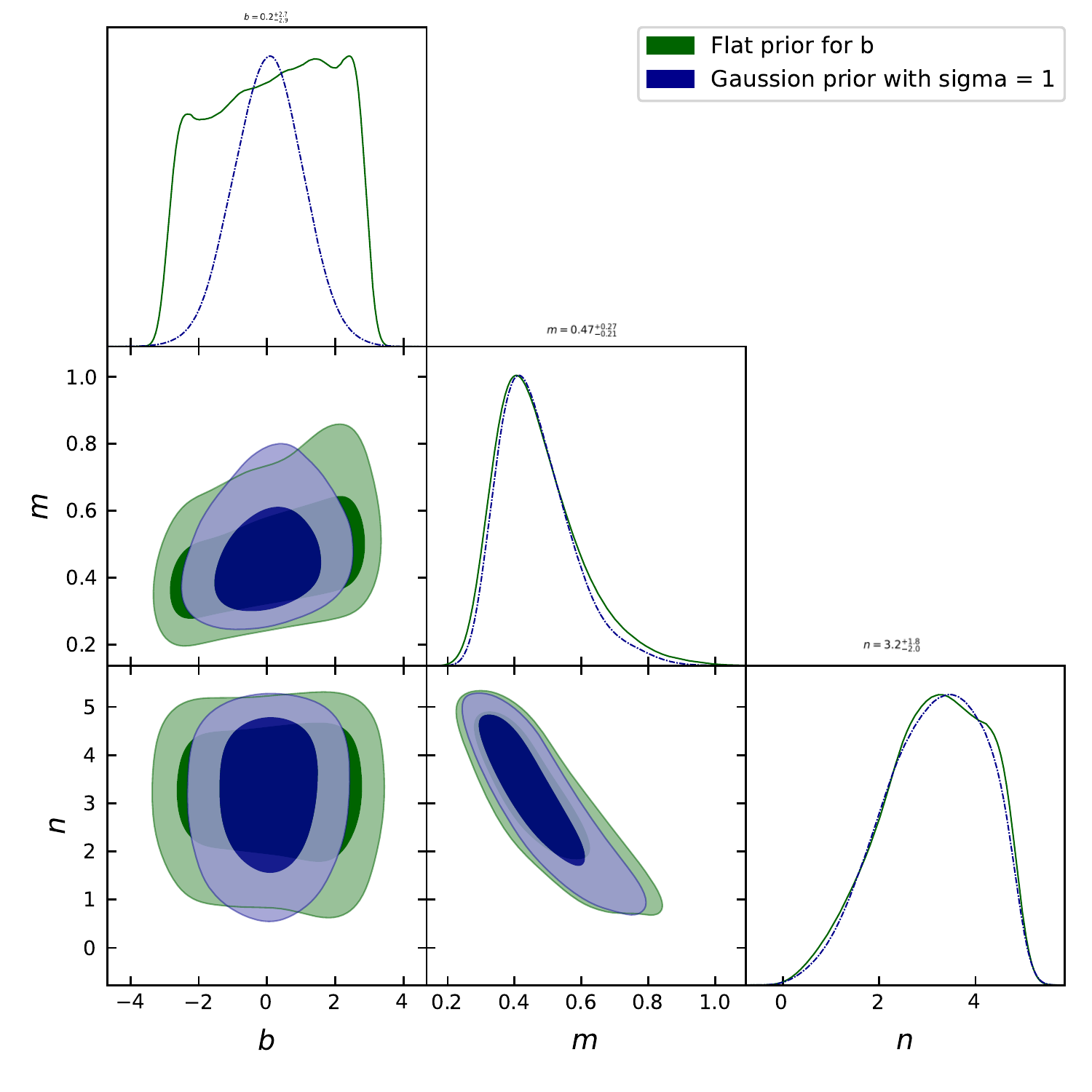}
\caption{The plot compares the two numerical analyses with different prior distributions for the parameter $b$. For either case, we considered a uniform distribution for the other two parameters, $m$, and $n$. The counter represent $1-\sigma$ and $ 2-\sigma $ confidence intervals.} \label{Fig-Lastfigure}
\end{figure*}

We present the results in the form of triangle plots in Fig. \ref{Fig-Lastfigure}, where, in addition to the parameter space, we also observe the marginalised distribution for the parameters $b$, $m$, and $n$ in our model.  The contour represent the $1-\sigma$ and $ 2-\sigma $ confidence intervals.

Initially, we perform the analysis considering a flat prior for all the parameters; however, we notice the marginalized distribution for parameter b to be roughly uniform in the range. We then motivate our work to study the result in the neighborhood of $b=0$. This approach intends to find any deviation from GR, which accounts for a local minimum for the function in Eq (\ref{21}). We also perform the numerical analysis with a Gaussian prior for the parameter $b$ with $\sigma=1.0$ as dispersion. The results are presented in Fig. \ref{Fig-Lastfigure}; there is no significant difference in the marginalized distributions of the remaining parameters, $m$, and $n$. 

\begin{figure*}[htbp]
\centering
\includegraphics[scale=1.5]{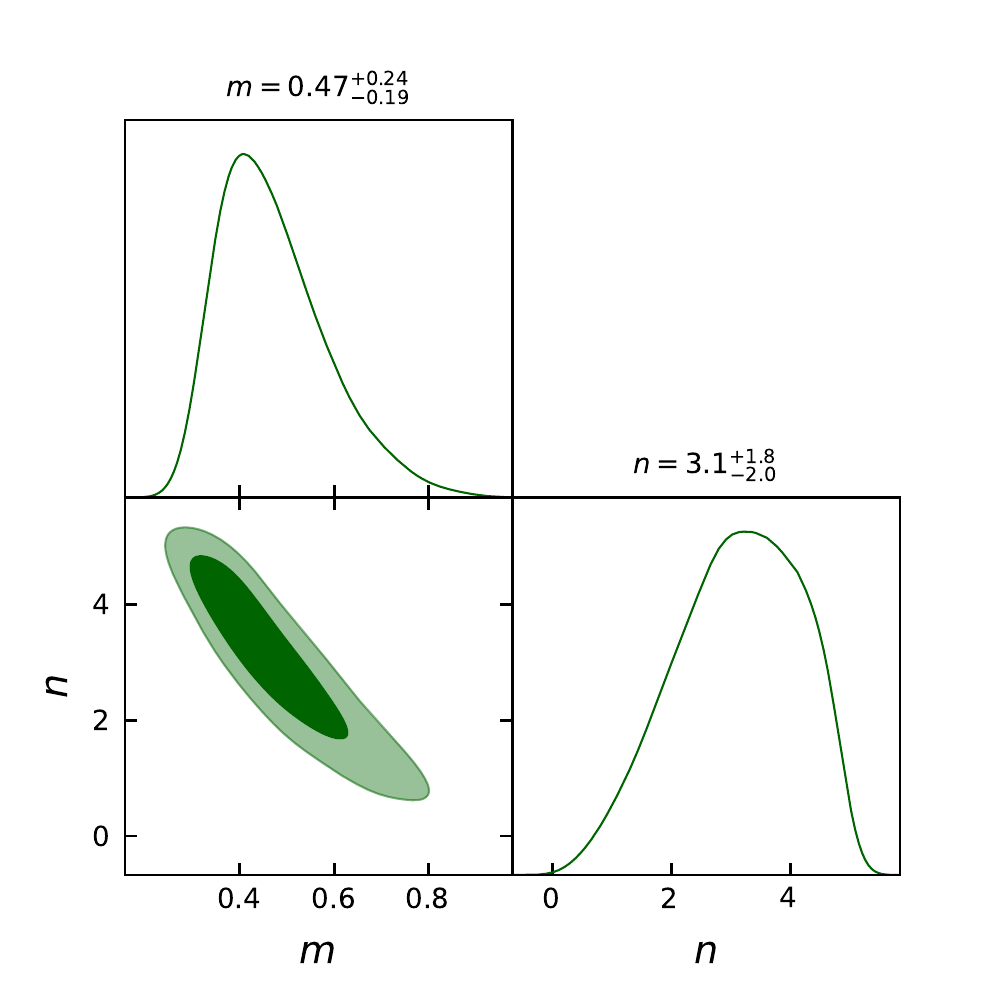}
\caption{The plot showing the best fit values of the model parameters $m$ and $n$ obtained with 1048 points of Pantheon datasets at $1-\sigma$ and $ 2-\sigma $ confidence level with $b=0$.} \label{Fig-mn}
\end{figure*}

If we consider the case of $b=0$, the model reduces to $f(Q,T)= f_{\Lambda}(Q)=Q$, i.e. it has a direct link to $\Lambda$CDM model. Therefore, the equation of Hubble parameter $H$ reduces to 
\begin{equation}
H(z)=H_{0}\left(\frac{m (z+1)^n+1}{m+1}\right)^{\frac{3}{2 n}}
\end{equation}

where $m$ and $n$ are model parameters. One interesting point regarding this expression of Hubble parameter is that for $n = 3$, this becomes exactly like the $\Lambda$CDM model as stated by \cite{Ankan/2016}.
The constraints for $m$ and $n$ using Pantheon SNeIa datasets is shown in Fig. \ref{Fig-mn}.

\section{Cosmological parameters} \label{sec5}

In this section, we study the behavior of the deceleration parameter and the equation of state parameter. 

\subsection{The deceleration parameter}
 
The deceleration parameter as a function of Hubble parameter $H$ is defined as 
\begin{equation}
q= -1- \frac{\dot{H}}{H^{2}} \,.
\end{equation}
In any cosmological model, the deceleration parameter $q$ play an important role in characterizing the decelerated phase($q>0$) and an accelerated phase($q<0$) of the universe. There are many works in literature in which deceleration parameter is useful in explaining the evolution phase of the universe \cite{Mamon,Sahoo}.
The equation of deceleration parameter $q$ according to our model reads,
\begin{equation} \label{26}
q = -1-\frac{3 (b +8 \pi ) m (-z-1) (z+1)^{n-1}}{b +(3 b +16 \pi ) \left(m (z+1)^n+1\right)}.
\end{equation}

\begin{figure}[H]
\centering
\includegraphics[width=8.5 cm]{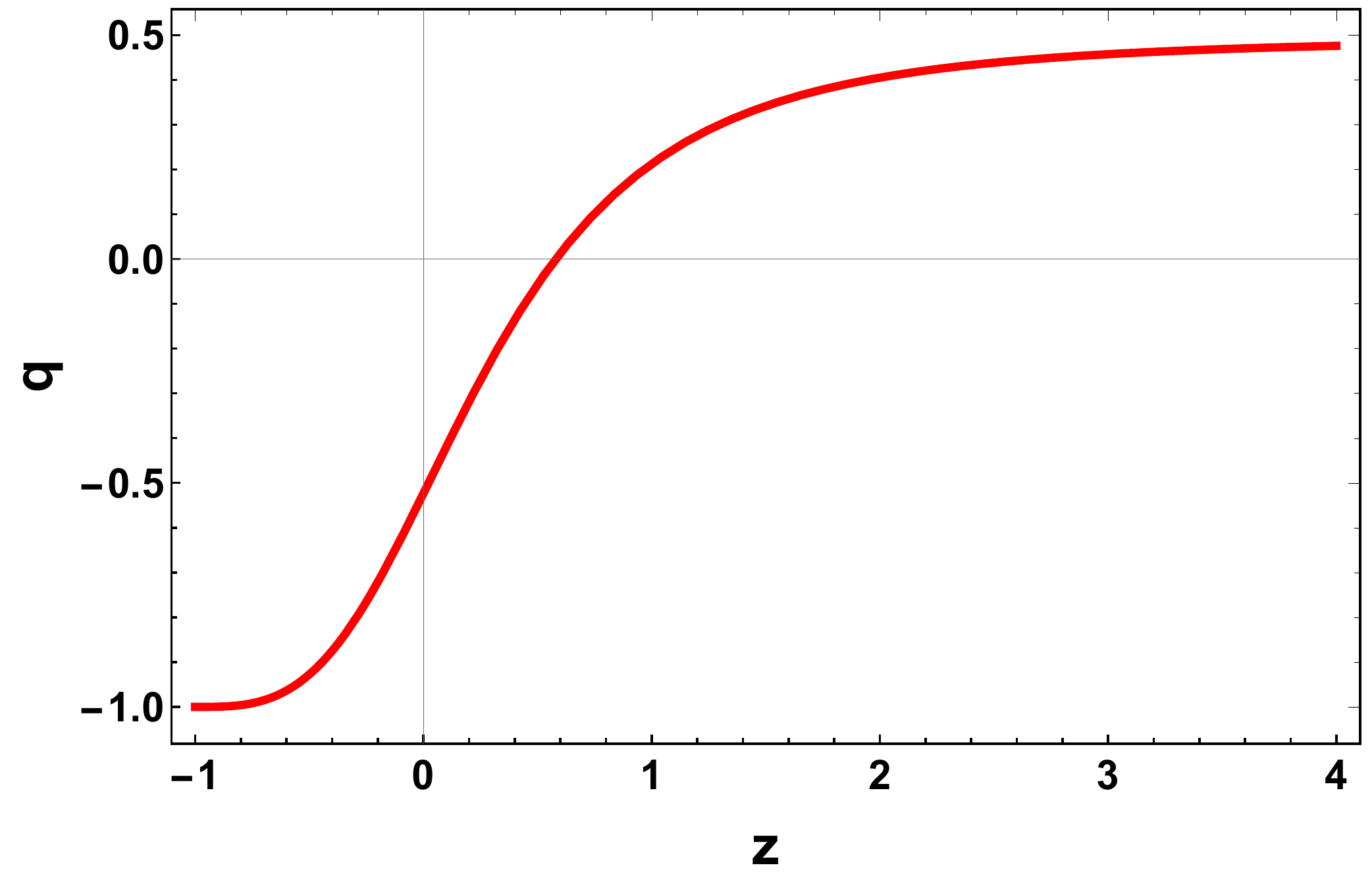}
\caption{Variation of deceleration parameter $q(z)$ for the best fitted value of $b=0.2$,  $m=0.47$ and $n=3.2$ from the analysis of SNeIa Pantheon samples.} \label{Fig-qz}
\end{figure}

The behavior of $q(z)$ is shown in Fig. \ref{Fig-qz} according to the estimated values of model parameters $b$, $m$ and $n$ by Pantheon sample. We can observe that there is a well behaved transition from deceleration to acceleration phase at redshift $z_{t}$. The value of the transition redshift is $z_{t}= 0.58\pm 0.30$ \cite{Farooq}. The result obtained is consistent with several works in literature \cite{Cunha,JV}. Also, we can note that the value of $q_{0}= -0.52$ \cite{Christine} which is negative at present indicating an acceleration in the universe.

\subsection{The EoS parameter}

The equation of state parameter as a whole determines a relation between energy density and pressure. The common phases observed through EoS parameter includes the dust phase at $\omega=0$. Then, $\omega=\frac{1}{3}$ shows the radiation-dominated phase whereas $\omega=-1$ corresponds to the vacuum energy i.e. $\Lambda$CDM model. Besides this, the accelerating phase of the universe which is in recent discussion is depicted when $\omega<-\frac{1}{3}$ which includes the quintessence ($-1<\omega\leq 0$) and phantom regime ($\omega<-1$). \\
In this work, we have considered and effective EoS parameter which contains two model parameters $m$ and $n$. 
According to the constrained values of m and n, the behavior of EoS parameter is shown below. The value of EoS parameter at $z=0$ is $\omega_{0}= -0.68^{+0.10}_{-0.11}$ \cite{Christine} which clearly indicates an accelerating phase.

\begin{figure}[H]
\centering
\includegraphics[width=8.5 cm]{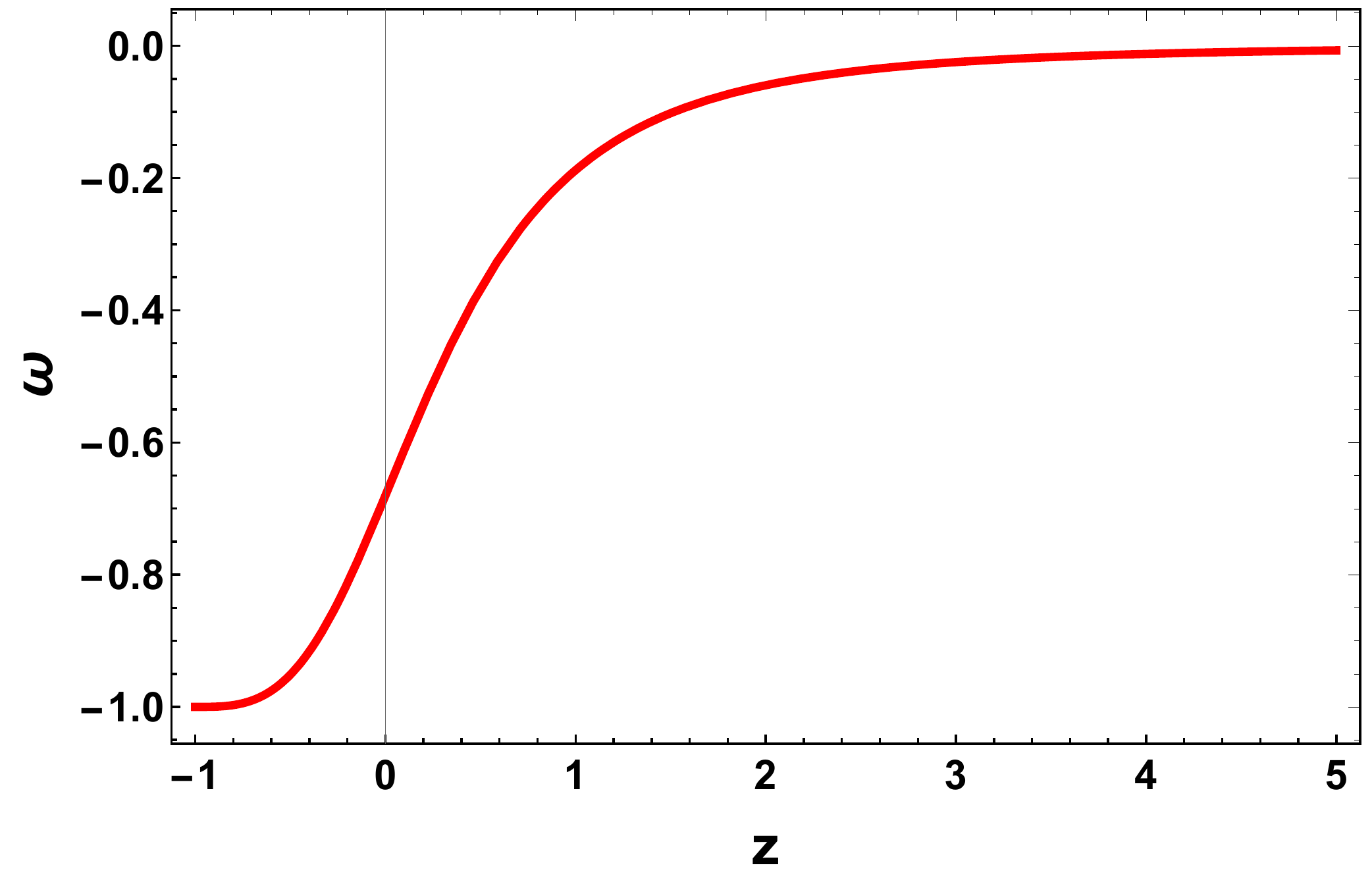}
\caption{Variation of EoS parameter for the best fitted value of $m=0.47$ and $n=3.2$ from the analysis of SNeIa Pantheon samples.} \label{Fig-Omega}
\end{figure}

\section{Conclusion}\label{sec6}

As new theories of gravity emerge in the literature, it is vital to put them to the test to see if they are viable in describing the dark sector of the universe. The $f(Q,T)$ is a promising new gravity theory based on a combination of the non-metricity function $Q$ and the trace of the energy-momentum $T$. To begin, we considered the functional form $f(Q,T) = Q+bT$, where $b$ is a free parameter.\\
We use a well-motivated parametric form of the equation of state parameter as a function of redshift $z$ to solve the field equations for $H$. It has a negative value of less than $-\frac{1}{3}$ at the epoch of recent acceleration. At a high redshift $z$, the value of $\omega$ tends to zero for positive values of the model parameters $m$ and $n$ and depends on the model parameter at $z = 0$. The Pantheon study, a recently proposed observational dataset, was used to constrain the parameter space. The parameters from our study have 2-$\sigma$ limits of $b = 0.2^{+2.7}_{-2.9} \ , m = 0.47^{+0.27}_{-0.21},\ n = 3.2^{+1.8}_{-2.0}$.
The error bar plot of the 1048 points in the Pantheon datasets and our obtained model compared to the $\Lambda$CDM model considering $\Omega_{m_{0}}= 0.3$, $\Omega_{\Lambda_{0}} = 0.7$ and $H_{0}= 67.4\pm 0.5km s^{-1}Mpc^{-1}$ shows a good match to the observational results.\\
Finally we observed the behavior of the deceleration parameter and equation of state parameter. We can see that there is a well behaved transition from deceleration to acceleration phase at redshift $z_{t}$. The value of the transition redshift $z_{t}= 0.58 \pm 0.30$ with the value of $q_{0}= -0.52$. And also, the value of EoS parameter at $z=0$ is $\omega_{0}= -0.68^{+0.10}_{-0.11}$ which clearly supports an accelerating phase. The present analysis motivates and encourages the study of such extensions, which may represent a geometric alternative to dark energy. Apart from the cosmic evidence, we also need to ensure that the new $f(Q, T)$ theory is stable so that the differences in approach from $\Lambda$CDM can be performed at the perturbation level. This can be investigated in detail for $f(Q, T)$ through future works.

\section*{Acknowledgments}

SA acknowledges CSIR, Govt. of India, New Delhi, for awarding Junior Research Fellowship. PKS acknowledges CSIR, New Delhi, India for financial
support to carry out the Research project [No.03(1454)/19/EMR-II Dt.02/08/2019].


\end{document}